# Effects of surface topography on low Reynolds number droplet/bubble flow through constricted passage

Aditya Singla and Bahni Ray[+]
Department of Mechanical Engineering, Indian Institute of Technology Delhi
New Delhi - 110016, India

[+]Author to whom correspondence should be addressed: bray@mech.iitd.ac.in

## Abstract
This paper is an attempt to study the effects of surface topography on the flow of a droplet (or a bubble) in a low Reynolds number flow regime. Multiphase flows through a constricted passage find many interesting applications in chemistry and biology. The main parameters which determine the flow properties such as flow rate and pressure drop, and govern the complex multiphase phenomena such as drop coalescence, break-up and snap-off in a straight channel flow are the viscosity ratio, droplet size and ratio of the viscous forces to the surface tension forces (denoted by Capillary number). But in flow through a constricted passage, in addition to the above-mentioned parameters, various other geometric parameters such as constriction ratio, length and shape of the constriction, phase angle, and spacing between the constrictions also start playing an important role. Most of the studies done on the problem of drop flow through a constricted passage have aimed to understand the role of physical parameters, with some studies extending their analysis to understand the variation of one or two geometric parameters. But no study could be found which explicitly evaluates the role of surface topography. An attempt has been made to unify the current literature as well as analyze the effect of the geometric parameters by understanding the physics and mechanisms involved. The non-dimensional numbers which govern this problem are then identified using the scaling analysis.

## 1. Introduction

Flow of emulsions through porous media finds many useful applications in oil and gas, food, chemical, pharmaceutical, and textile industries [1-8]. To simplify the problem of emulsion flow through porous media, many researchers have treated this problem analogous to drop/bubble flow in a constricted channel. Multiphase flows through a constriction also find use in many biological applications such as detection of cancer cells in the blood [9] and study of mechanical properties of red blood cells and three-dimensional protein networks in the blood [10]. These applications discussed above usually involve low Reynolds number (the ratio of inertial forces to the viscous forces, denoted by *Re*) flows and hence the inertial effects are not so dominant, and various surface and capillary effects start coming into picture.

A very interesting and important field that has recently developed with low *Re* flows as its foundation is microfluidics, which involves flows through micrometer dimensions. Since the characteristic length (and in most cases, the velocity too) is very small (order of 10-100 µm), the Reynolds number is very low. The flow physics changes drastically and becomes a lot more interesting at the microscale, owing to the various surface effects that start coming into picture such as surface tension, Marangoni effect, electric effects, etc. Microfluidic devices involving droplets and bubbles are now being used in various "Lab-on-a-chip" applications in which various chemical, biological processes and reactions are performed conveniently at the microscale [10-14], leading to exceptional progress in biomedical technology in the recent years. Another advantage of microfluidic flows is the ability to produce polydisperse droplets a lot more easily. Various methods have been developed for generating a continuous stream of microfluidic droplets [15-17]. These can be broadly categorized into three categories: droplet generation using co-flowing streams [18,19], droplet generation using cross-flowing streams in T-junction [20,21] and droplet generation using elongational flows (flow-focusing techniques) [22,23]. A particular term has been coined for this study of droplets at microscale- "Droplet Microfluidics". It is essentially the study of low *Re* flows by taking various surface effects into account. As was mentioned very aptly by H.A. Stone [24] in his review on microfluidic devices - "In particular, the most important issues are not "macro" versus "micro" but rather the relative magnitude of various effects as characterized traditionally using dimensionless parameters." Stone [24] also provided a list of various geometric (network connectivity, channel cross section and curvature, surface topography and porosity), chemical (wettability, surface charge, chemical affinity and ionic strength) and mechanical (hardness, elasticity) characteristics.

In this review paper, we describe various works on the effects of surface topography of a constricted passage on droplet/bubble deformation. To understand what factors affect the drop flow through a constricted passage, it is essential to first understand the parameters which affect the flow in a straight channel, as those factors would anyways play a role in the former case too. This has been shown in Section 2. Following that, the surface topographic effects, which play an important role in determining the flow properties and bubble morphology, have been studied in section 3. In section 4, a new scaling analysis has been proposed for flow of droplet through constricted passage with any surface topography. Finally, in section 5, the different applications of above phenomena are discussed.

## 2. Flow through a straight channel

Analysis of droplet flow in a channel has been a topic of interest for long. The main factors which affect the multiphase phenomena and the flow properties (velocity profile and pressure drop) are the viscosity ratio ($\lambda$), the droplet size (*a*) and the Capillary Number (*Ca*). The viscosity ratio is the ratio of droplet viscosity ($\mu_2$) to the surrounding fluid viscosity ($\mu_1$), i.e. $\lambda = \mu_2/\mu_1$. For a bubble, the viscosity ratio tends to zero. The droplet size is generally taken relative to the channel radius and is defined as the ratio of undeformed drop radius to the tube radius. The most important parameter is Capillary number, which measures the relative significance of viscous forces to the interfacial forces, the two major forces in low *Re* flow. Thus, the capillary number is defined as $Ca = \mu_1 V/\sigma$, where *V* is the mean velocity of the surrounding fluid, and $\sigma$ is the interfacial surface tension. The droplet and the surrounding fluid are assumed to be immiscible.

Initial studies in this area were focused on calculating the bubble velocity, the additional pressure drop across the bubble and the film thickness for different bubble fluids and surrounding fluids [25-30]. The extra or the additional pressure drop is defined as the pressure drop over a distance extending into the undisturbed flow ahead and behind of the drop, minus the pressure drop in the normal Poiseuille flow over the same distance.

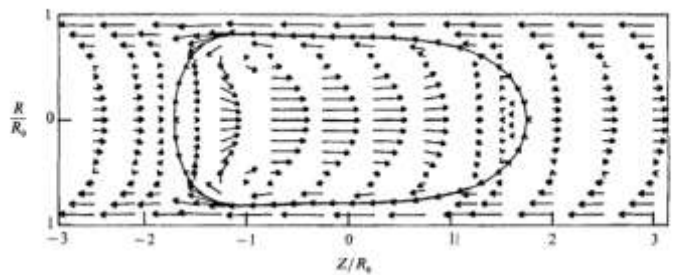

Figure 1: Velocity vector field in and around a drop for *a* = 1.10, *Ca* = 0.10, $\lambda$ = 0.19. Velocities shown are relative to the drop velocity. Reproduced with permission from Journal of Fluid Mechanics 210, 565-591 (1990). Copyright 1990, Cambridge University Press.

Ho and Leal [31] did experimental studies to evaluate the extra pressure drop, velocity and shape of the drops for different viscosity ratios, capillary numbers and drop sizes by taking many different Newtonian as well as viscoelastic systems. Shen and Udell [32] evaluated velocities, pressure drop and film thickness in bubble flow using the finite-element method. Martinez and Udell [33] studied flow of long bubbles in capillaries using the boundary integral method and evaluated bubble velocities and profiles. They continued their work to analyze axisymmetric motion of drops in circular tubes and computed drop speed and pressure drop for a wide range of capillary numbers, viscosity ratios and drop size [34]. Their work aptly summarizes and explains the effect of droplet flow in a straight channel (Fig. 1 and 2).

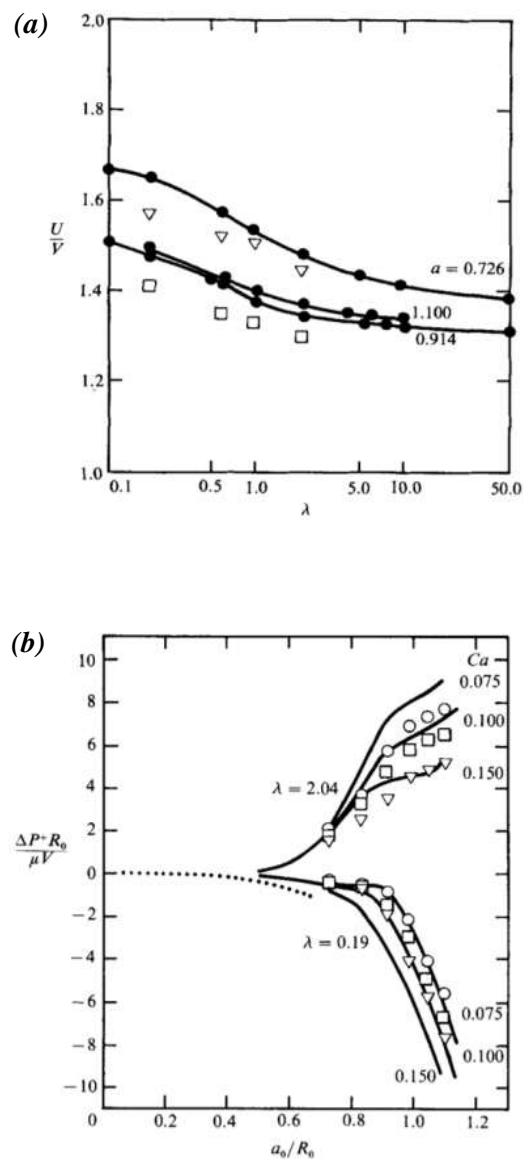

Figure 2: (a) Variation of drop velocity with viscosity ratio for different drop sizes; (b) Variation of additional pressure drop with drop size for different capillary numbers. Reproduced with permission from Journal of Fluid Mechanics 210, 565-591 (1990). Copyright 1990, Cambridge University Press

The result of variation of each of the three major parameters has been described below:

## 2.1. Effect of capillary number

Capillary number gives an idea of the relative importance of the viscous forces being exerted on the drop by the surrounding fluid to the interfacial forces. If capillary number tends to zero, it implies that the viscous forces are negligible and hence, the droplet acquires a perfectly spherical shape and the surface tension balances the excess pressure gradient between the inside and outside of the drop. This pressure difference for a spherical droplet is given by the Young-Laplace equation, $\Delta p = 2\sigma/A$, where $A$ is the radius of the spherical drop in dimensional units.

If the capillary number increases, the deformation of the droplet increases (due to greater viscous forces) and hence the film thickness, which is the thickness of the fluid surrounding the droplet, increases. Also, since the drop is deformed more with an increase in $Ca$, the drop velocity also increases due to the drop being localized about the centerline, where the fluid velocities are higher. The additional pressure drop decreases with an increase in $Ca$ [Fig. 2 (b)], but is also strongly dependent on the other two factors discussed below.

## 2.2. Effect of viscosity ratio

Viscosity ratio is the ratio of the droplet viscosity to the surrounding fluid viscosity. A larger viscosity ratio implies more resistance by the drop and hence a greater deformation. Also, greater resistance implies lesser velocities, hence the drop velocity decreases with the viscosity ratio [Fig. 2 (a)]. The effect of viscosity ratio on the pressure drop is very interesting. It has been observed that for viscosity ratios greater than 1, the additional pressure drop is generally positive, whereas for viscosity ratios lesser than 1, it is negative [Fig. 2 (b)]. The viscosity ratio, $\lambda = 1$ corresponds to droplet viscosity equal to the surrounding fluid viscosity and can be thought of as producing negligible additional pressure drop. Of course, this is a lot dependent on capillary number and the drop size. For example, for extremely small capillary numbers, the additional pressure drop is positive even for viscosity ratios less than 1.

## 2.3. Effect of drop size

The drop size is generally given with respect to the tube radius. A larger drop undergoes a greater deformation. The film thickness decreases with increase in drop size due to the bulkiness of the drop, so does the drop velocity [Fig. 2 (a)]. It has also been found that the magnitude of additional pressure drop increases with the size of the drop, for both viscosity ratios lower and greater than 1 [Fig. 2 (b)].

Olbricht and Kung [35] demonstrated that break-up of drops may take place even in a straight channel if the capillary number is too large. They termed the capillary number at which this process starts for a given drop size and viscosity ratio as the critical capillary number. They inferred that critical capillary number increases with drop size and decreases with viscosity ratio.

## 3. Flow through channel with surface topography

Changing the surface geometry of the channel can lead to substantial changes in drop/bubble deformation and flow properties. The first study on droplet flow through a constricted passage was done by Olbricht and Leal [36], where they experimentally studied how the Capillary number, drop size and viscosity ratio affect the flow parameters such as drop velocity and the pressure drop as well as the drop morphology inside a constricted sinusoidal tube.

One of the two main striking features which they found out for a constricted channel flow was that when the drop passes through the constriction, the pressure drop across the channel changes considerably and approaches a sinusoidal distribution for large droplet sizes (Fig. 3). Though the average values may match the value in flow through a straight channel, the peak values of pressure (also called peak resistance), which occur when the drop is flowing through the narrowest part of the constriction, are significantly higher than the average value. Under conditions of imposed pressure gradient, extremely high peak resistances may lead to drop being retained at certain locations in the channel. Hence, the geometry needs to be modeled very carefully to get the desired results.

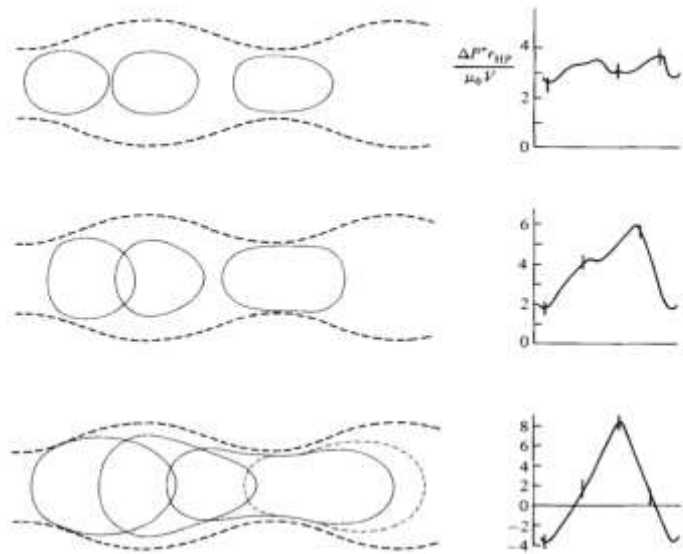

Figure 3: Variation in drop shape and additional pressure drop in flow through a constricted channel for increasing drop sizes (from top to bottom), $Ca = 0.043$. Additional pressure drop approaches a sinusoidal profile as the drop size increases. Reproduced with permission from Journal of Fluid Mechanics 134, 329-355 (1983). Copyright 1983, Cambridge University Press

The other important feature observed was that for the same parameters as a straight channel, drop break-up was observed in a constricted channel. The breaking-up of drop can be understood by analyzing the viscous forces and the surface tension forces when break-up occurs. Depending on the capillary number, the break-up may take place by different mechanisms such as dripping and jetting. An increase in capillary number leads to a transition from a dripping regime to jetting regime.
To understand these mechanisms, we consider the case of droplet formation via coflowing liquid streams where the dispersed phase channel is enclosed inside the continuous channel or the surrounding fluid channel. At low flow rates, periodic drop formation takes place by dripping mechanism where the liquid pinch-off occurs at the tip of the channel. The viscous drag on the drop by the surrounding fluid scales as $\mu_1 V d_{drop}$, where $d_{drop}$ is the drop diameter (in dimensional units), $\mu_1$ is the surrounding fluid viscosity and $V$ is the mean velocity of the surrounding fluid. The drop also experiences a surface tension force which tries to hold the drop. This force scales as $\sigma d_{tip}$, where $d_{tip}$ is the diameter of the tip where the break-up occurs and $\sigma$ is the surface tension. As the drop radius increases, the viscous drag increases because of the increase in drop size. The diameter of the breaking drop can be calculated by balancing the two forces as the surface tension will be unable to hold the drop for a further increase in viscous drag [19]. Thus, $\mu_1 V d_{drop} \sim \sigma d_{tip}$, which gives $d_{drop}/d_{tip} \sim \sigma/\mu_1 V$ or $d_{drop}/d_{tip} \sim 1/Ca$. A more accurate relation is obtained by equating the drag force obtained from Stokes formula with the surface tension forces, which yields $d_{drop}/d_{tip} \approx 1 + 1/3Ca$ [18]. Therefore, the drop size decreases with the increase in capillary number and a critical diameter is reached after which the viscous forces start stretching the dispersed phase to form a jet. The break-up will now be influenced by the Rayleigh-Plateau instability also, in addition to the viscous drag.

To get a clearer picture of the transition from dripping to jetting regime, we consider the case of drop generation by flow-focusing or elongational flow [16]. In such a flow, the dispersed phase is elongated by the surrounding fluid and then made to pass through a narrower area. Dripping corresponds to break-up taking place right after the constricted area whereas in jetting, a jet formation takes place after the constricted area. Since the elongation is taking place for both regimes now, the break-up occurs by the combined effect of viscous drag and capillary instability (which includes the Rayleigh- Plateau instability and end-pinching). Therefore, the diameter of drop in this regime cannot be simply computed by force balance. The transition from dripping to jetting regime occurs when the capillary number increases beyond a critical capillary number and can be understood in terms of the timescales involved in this problem. The timescale for pinch-off ($t_{pinch}$) is given as $d_{jet}\mu_1/\sigma$, where $d_{jet}$ is the jet diameter, and the timescale for jet growth ($t_{growth}$) is given as $d_{jet}^3/Q_d$, where $Q_d$ is the flow rate of the dispersed phase. The capillary number can be defined as the ratio of these two timescales. The critical capillary number for which the transition from dripping to jetting takes space is given when the two timescales are comparable, which gives $Ca_{critical} = t_{pinch}/t_{growth} = Q_d\mu_1/\sigma d_{jet}^2 \sim 1$. So, for $Ca < Ca_{critical}$, the interface instability is able to grow at the outlet and hence the dripping regime is observed, whereas for $Ca > Ca_{critical}$, the growth of jet occurs faster and is able to suppress the instability, resulting in a jetting regime (the drops may later be formed in the jetting regime too as the jet thickness increases). In the flow of a drop through a constricted passage too, the break-up is governed by the effects of capillary instability and viscous drag, which has been discussed in subsequent sections.

Lastly, a constriction may also promote drop coalescence by allowing two parallel streams flowing through a constriction, which may usually not be possible in a straight channel. Yan et al. [37] studied this phenomenon and observed that this occurs at relatively lower capillary numbers since at higher capillary numbers, the enhanced deformation allows the drop to move through the constriction without undergoing coalescence.

Many other researchers have analyzed the effect of the above-mentioned parameters (*Ca*, *a*, *λ*) as well as the constriction parameters on the drop flow. Some reviews on flow through porous media [2, 38] have also been done, which give the reader a general idea of how a constriction affects the flow. We try to understand the role of surface topography by looking at five major geometry parameters:

### 3.1. Effect of constriction shape:

The shape of the constriction plays a substantial role in determining the drop morphology and can be altered in many ways. The shape may be semi-circular, rectangular, trapezoidal, sinusoidal etc. The most common shape that has been studied is perhaps the sinusoidal geometry. The common form of the channel shape for flow through a sinusoidal converging/diverging tube is given by:

$$r(z) = r_a + \frac{d}{2} \sin(2\pi z / L) \qquad (1)$$

where *z* is along the flow direction, $r_a$ is the average radius of the tube, *L* is the length of the constriction and *d* is the depth of the constriction (maximum radius, minus the minimum radius). The maximum radius is the tube radius, which is denoted *R*. by These parameters can be defined in a similar way for any kind of shape.

In addition to the study by Olbricht and Leal (discussed in previous section), a lot of other studies have also been done on drop flow through **sinusoidal constrictions,** including Tsai and Miksis [39] who analyzed this problem in detail using the boundary integral method. They also studied the effect of physical parameters on drop break-up. If the capillary number is too less, there is less fluid around the drop when it passes through the constriction (because of thinner film due to lesser deformation), so it takes longer for the film instability to develop. As the capillary number is increased, this instability is increased and snap-off starts taking place. But if the capillary number is too high, the snap-off doesn't take place, because of the large velocities which do not give time for instabilities to grow (though there is still a very minute portion of fluid which breaks off). Hence, there is a range of capillary numbers for which the breaking occurs. This was also evident from the results as the snap-off was observed for $Ca = 0.1$ but not for $Ca = 0.05$ and $Ca = 0.2$ (for $a = 0.9$, $\lambda = 0.001$). These values vary with change in drop size. A larger drop gives more time for the instability to grow and the snap-off now starts taking place at lower capillary numbers also. For $a = 1.2$ and $\lambda = 0.001$, the snap-off was observed even for $Ca = 0.05$, in addition to $Ca = 0.1$, but for $Ca = 0.2$, the break-up still didn't occur [Fig. 4 (a)-(c)]. It was also observed that lowering the viscosity ratio promotes snap-off since less viscous drops cannot support high velocity of the surrounding fluid. A lower velocity of the surrounding fluid implies more time for the instability to grow, which results in snap-off. For $Ca = 0.1$, $a = 0.9$, snap-off was observed for $\lambda = 0.0005$ but not for $\lambda = 0.01$.

From the above reasoning, it can also be concluded that for the range of capillary numbers for which the snap-off occurs, the increase in capillary number supports the drop break-up and therefore the snap-off time ($\tau$) decreases with increase in capillary number (the snap-off time is the difference between the time when the drop passes the constriction center and the time when snap-off occurs). Since the snap-off time decreases with increasing $Ca$, this means that the radius of the drop which breaks apart, $b_r$, will be smaller when capillary number is increased [Fig. 4 (d) and (e)].

*(a)*

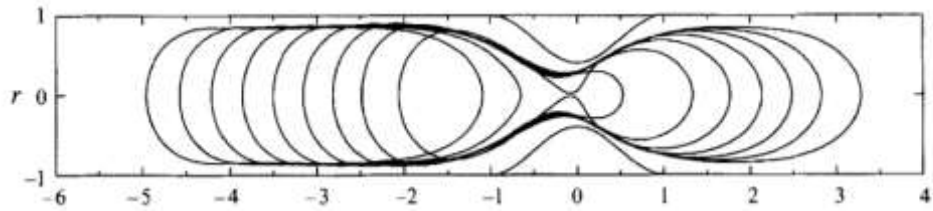

*(b)*

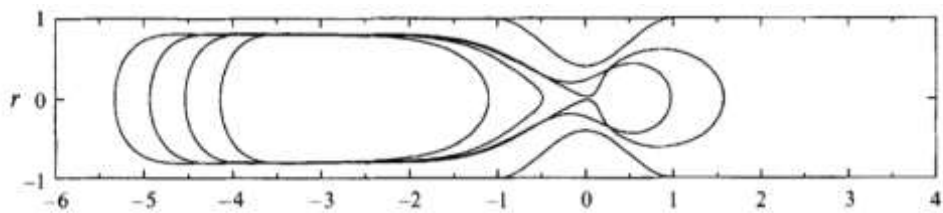

*(c)*

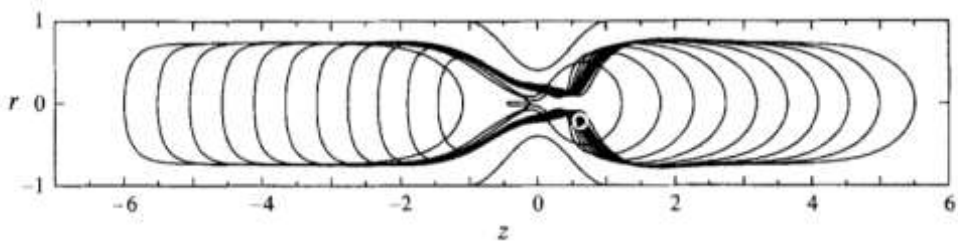

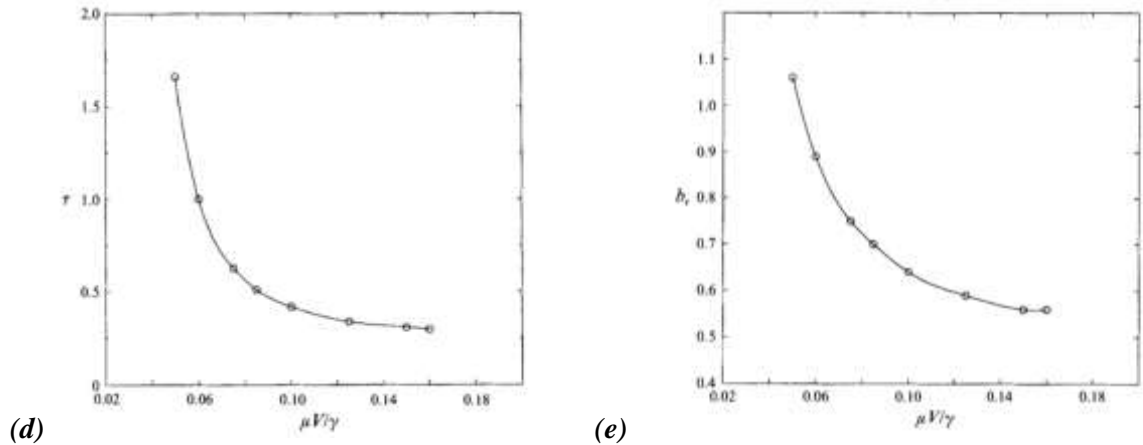

*(d)*   *(e)*

Figure 4: Effect of capillary number on snap-off of a drop in a sinusoidal constriction, $a = 1.2$, $\lambda = 0.001$; (a) $Ca = 0.05$; (b) $Ca = 0.1$; (c) $Ca = 0.2$; (d) Effect of $Ca$ on snap-off time ($\tau$) for $a = 1.2$, $\lambda = 0.001$; (e) Effect of $Ca$ on radius of breaking drop radius ($b_r$) for $a = 1.2$, $\lambda = 0.001$. Reproduced with permission from Journal of Fluid Mechanics 274, 197-217 (1994). Copyright 1994, Cambridge University press.

Another type of constriction shape which has been studied includes **rectangular geometry**. Nath et al. [9] analyzed motion, deformation and behavioral changes of EMT cancer cells passing through rectangular micro capillaries to know the exact mechanism of cancer spreading, which is not known precisely yet. They studied how the cancer cells invade the blood capillaries via basement membrane and how the motion of cells through the blood micro capillaries takes place. For the former case, they had constrictions having gaps in the range of 10–15 µm and for the second case, the gap was 7 µm. They calculated the deformation index, entry time and velocity of the cells for different gap sizes and cell sizes. Such parameters are important to understand the ability of cancer cells to move through the constricted capillaries. In this case, cells can be considered equivalent to droplets as they too undergo deformation and can be modeled as different phase than that of surrounding fluid i.e. blood. It was found that the cells are deformed to a great extent inside the micro channel, which is still prominent after they come out of the constriction (Fig. 5). This property of having enhanced deformation can be used to study the elastic properties of protein networks in the blood as was mentioned in one of the applications in the beginning [10]. It should also be noted that it is not necessary that drop break-up will not happen in rectangular constrictions since it depends on a lot of other parameters like capillary number, as was seen earlier. A very large capillary number may indeed lead to a break-up.

*(a)* 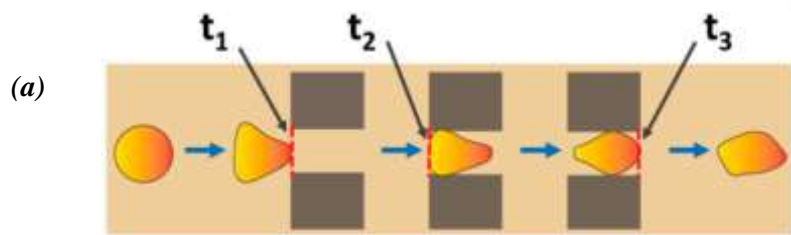

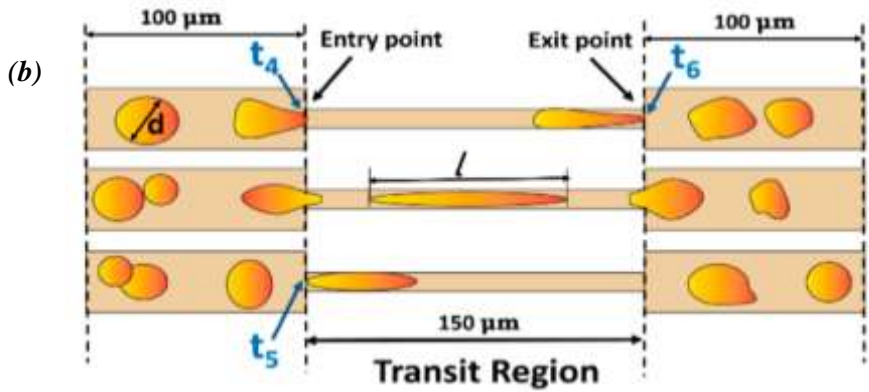

Figure 5: (a) Deformation and elongation of cells in rectangular constrictions resembling the basement membrane. Gap size is in the range of 10–15 µm; (b) Deformation and elongation in constrictions resembling microcapillaries. Gap size is 7 µm. Reproduced from J. Clin. Med. 8,1194 (2019). Copyright 2019, Multidisciplinary Digital Publishing Institute.

In addition to elongation, some interesting shapes have also been observed through a rectangular constriction for a smaller drop size and higher velocities. Chen et al. [40] observed crescent moon shaped bubbles coming out of a rectangular constriction [Fig. 6 (a)]. The reason for such a shape was that as the bubble approaches the constriction, the flow converges, and the bubble attains a higher velocity due to its small size and is elongated in the flow direction. After passing through the constriction, the velocity of the front part starts reducing in the diverging section. The back of the bubble is still travelling at a higher velocity, giving the crescent moon shape. They also quantified the bubble shape by using three parameters, namely, roundness, elongation, and convexity [Fig. 6 (b)]. As the bubble changes its shape from spherical to crescent-moon, the roundness and convexity clearly decrease. For a short period while passing through the orifice, the roundness increases as the bubble starts attaining a spherical shape again. The elongation or the elongation ratio gives an idea about the increase in length in the direction of the flow and it increases as the bubble starts approaching the orifice but starts to decrease as the bubble flows through the constriction.

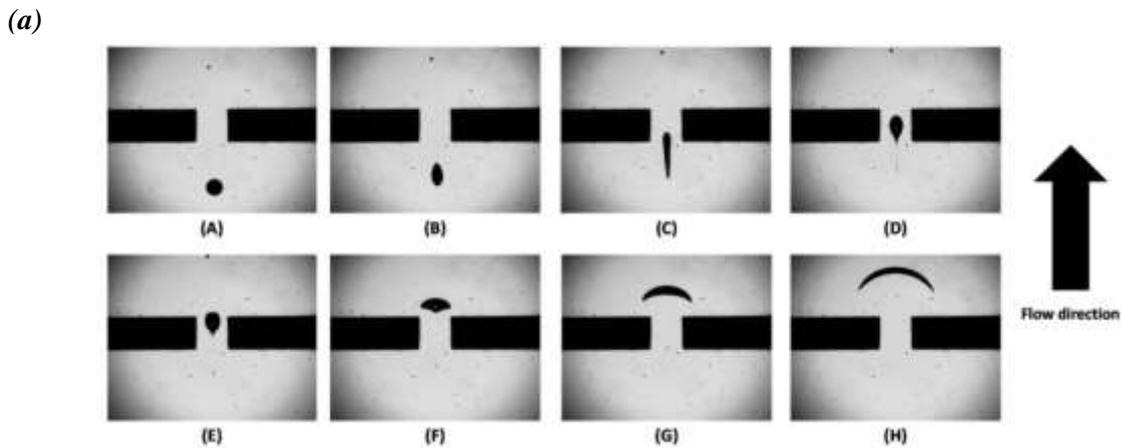

*(b)*

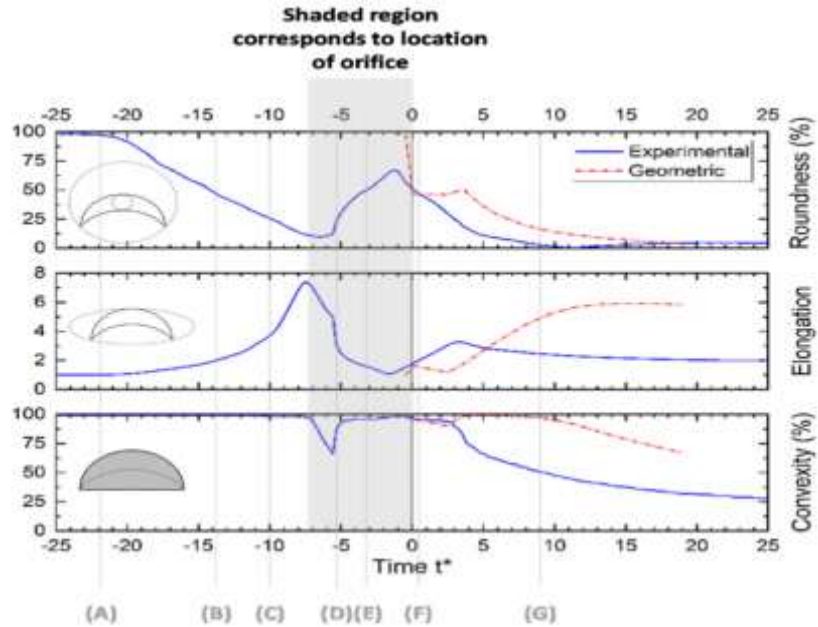

Figure 6: (a) Evolution of a spherical bubble of initial radius 745 µm passing through an orifice. Continuous phase flow velocity = 7.85 mm$^3$/s and (b) Plots depicting the roundness, convexity and elongation of the bubble as it passes through the constriction. Reprinted from Chemical Engineering Science, 206, C.-H. Chen, B. Hallmark, and J. F. Davidson, "The motion and shape of a bubble in highly viscous liquid flowing through an orifice", 224-234, Copyright (2019), with permission from Elsevier

Drop flow has also been studied through a constriction resembling a **trapezoid**. Cunha et al. [41] evaluated the mobility reduction factor and drop shapes for flow in such a geometry. The mobility reduction factor is a representative of the pressure and drop flow-rate relation in the flow and increases with decrease in maximum additional pressure drop. The drop shapes shown in the paper resemble those in the previous case of rectangular geometry (Fig. 5). This is because base length is taken to be very large due to which the drop is elongated and hence the trapezoid legs seem to play no significant role. It can be studied further till what ratio of leg length to base length will such a deformation be observed and when will snap-off take place. It is also observed that the drops elongate in the flow direction before the constriction and in the orthogonal direction after the constriction. Cabral and Hudson [42] also observed flow in trapezoidal-type geometry and a similar evolution of drop shapes can be seen. An important phenomenon that has been observed in similar geometries is drop coalescence, where the drop is stopped at the constriction, allowing the succeeding drop to come and merge with it [10, 43]. This phenomenon has also been Section 3.5.

Zinchenko and Davis [44] studied flow in a porous media using the boundary integral method where the porous media structure was modeled using either spheres or disks. For example, an arrangement of two spheres, placed at a certain distance, is used and the drop passes symmetrically through them. This can thus be viewed as flow through a **semi-circular constriction.** The drop deforms perpendicular to flow direction as expected but some interesting shapes are obtained if the results are analyzed in 3-D. It is seen that the drop surface starts becoming dimpled [Fig. 7 (a) and (b)], which would have not been visible, had the drop been seen in 2-D. The dimple starts to disappear when the drop comes out of the constriction and the drop takes the form of an elongated sphere, and ultimately returns to a spherical shape further downstream. The drop velocity with time is shown in Fig. 7-(c) for $Ca$ = 0.63, 0.45, 0.36 and 0.225 ($a$ = 0.9, $\lambda$ = 4). The constriction slows down the droplet and hence the velocity attains a minimum inside the constriction for *Ca* = 0.63, 0.45 and 0.36. But this is not the case for *Ca* = 0.225. This is because for lower capillary numbers, the viscous forces which govern the motion of droplet are unable to overcome the pressure resistance offered by the constriction. Hence, the drop becomes trapped and reaches a steady state without being able to pass through the constriction. More details on trapping of drops have been discussed in Section 3.2. Some other studies on drop flow in semi-circular constrictions have also been discussed in sections 3.2 and 3.5.

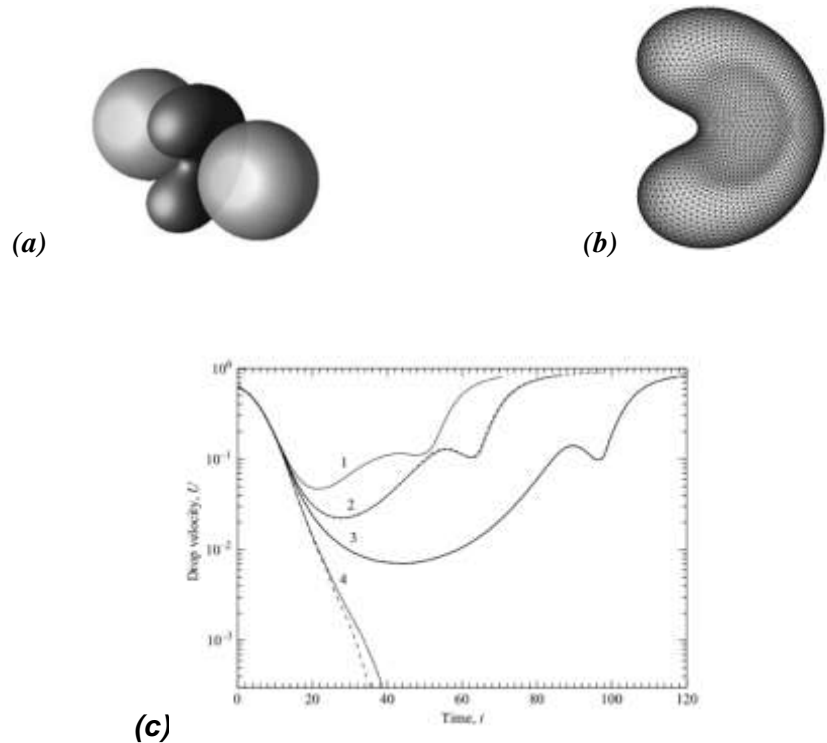

Figure 7: (a) and (b) Evolution of drop shape while passing through a two-sphere constriction for *Ca* = 0.63, *a* = 0.9, *λ* = 4; (c) Variation of drop velocity with time as the drop passes through the constriction for (1) *Ca* = 0.63, (2) *Ca* = 0.45, (3) *Ca* = 0.36 and (4) *Ca* = 0.225 (*a* = 0.9, *λ* = 4 for all cases). Reproduced with permission from Journal of Fluid Mechanics 564, 227-266 (2006). Copyright 2006, Cambridge University press.

Numerical studies of drop deformation have also been done in a **hyperbolic constriction** by Khayat et al. [45]. They calculated the deformation parameter, *α*, which is defined as the ratio of difference in current perimeter and the initial perimeter of the drop, to its initial perimeter. They observed that the deformation was symmetrical before and after the constriction, and *α* obtained a peak at the center of the constriction, thus corresponding to maximum deformation. This was for the case of a drop placed symmetrically initially. Droplets placed off-axis initially, in addition to undergoing rotation, also retain deformation while coming out of the constriction (Fig. 8), unlike the symmetrical case. Study on drop flow on a hyperbolic constriction was also done by Leyrat-Maurin and Barthe-Biesel [46], who analyzed the effects of capillary number and drop size on parameters such as drop shape and flow rate.

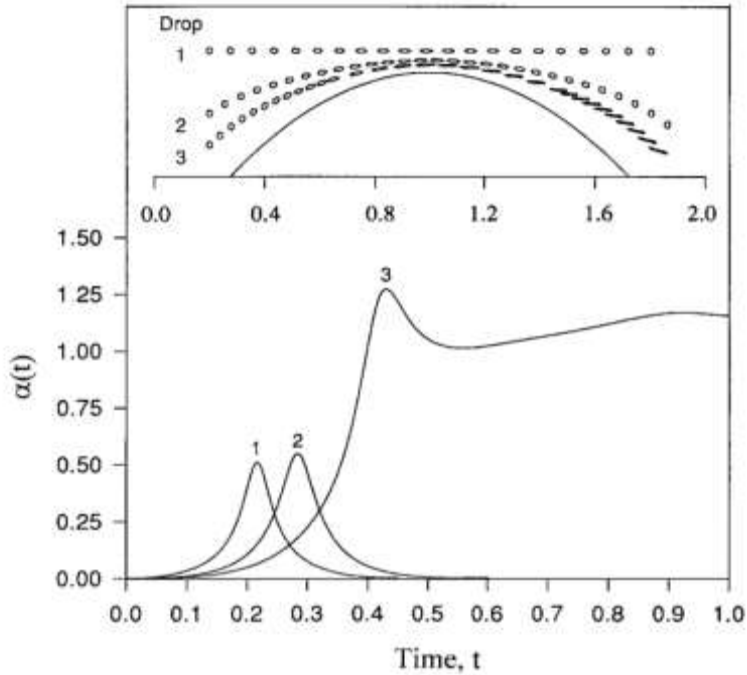

Figure 8: Drop shapes and deformation for symmetrical and unsymmetrical initial positioning of drops, $a = 0.02$, $\lambda = 4$. The droplets placed symmetrically (marked by 1) retain their shape, while the droplets placed off-axis (3) have high deformations on coming out of the constriction. Reprinted from International Journal of Multiphase Flow, 26, R. E. Khayat, A. Luciani, L. A. Utracki, F. Godbille, and J. Picot, "Influence of shear and elongation on drop deformation in convergent-divergent flows", 17-44, Copyright (2000), with permission from Elsevier

Roca and Carvalho [47] analyzed drop flow through a geometry consisting of **three tangent arcs of circles** and showed that the mobility reduction factor decreases with decrease in capillary number and increase in viscosity ratio and drop size. The drop shapes seemed to resemble those in sinusoidal geometries because of shape similarity.

Though it has been seen that different shapes of geometries can lead to different properties and drop shapes, the effects of a given shape may not be unique to it. It is just that such results have been observed for that shape for a given set of parameters, which may differ in different papers. While this is true, it is also true that certain constriction shapes are more suited to particular applications, as has been discussed above. To further elaborate this, we present some quantitative analysis below.

As we will see ahead, all other geometric parameters can be quantified in terms of a single numerical parameter. We wish to do the same for the constriction shape and hence define a shape parameter, $k$, as difference of the maximum area the constriction can have and the area of the constriction. The maximum area a constriction can have is (with given length and depth) that corresponding to a rectangular constriction, which is $Ld$. Therefore, $k$ can be written as:

$$k = Ld - A_c \qquad (2)$$

where, $A_c$ is the area of a single constriction.

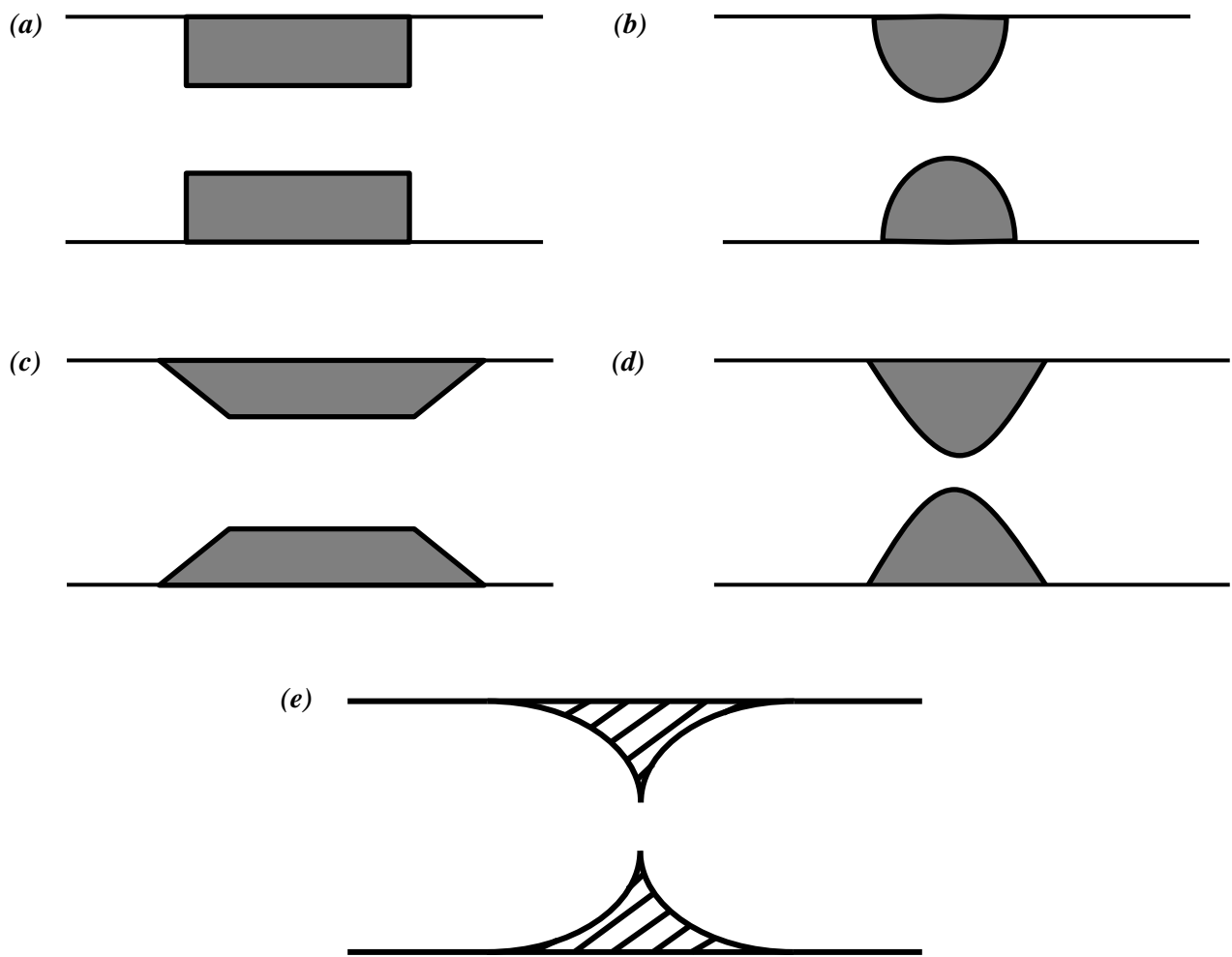

Figure 9: Schematic illustrating four different constriction shapes: (a) rectangular; (b) semi-circular; (c) trapezoidal; (d) sinusoidal; and (e) curved triangular geometry.

We will non-dimensionalize $k$ with $Ld$ and denote it by $\bar{k}$, where $\bar{k} = 1 - A_c/Ld$. This parameter has been defined in such a way that it captures the area reduction for a given shape. For example, there is no area reduction for a rectangular constriction (maximum area) and thus the shape parameter is 0. The change in shape can be analyzed in terms of change of area, given the same length and depth. Fig. 9 shows different shapes of constriction geometry and Table 1 shows the value of shape parameter for different shapes.

The effect of shape parameter on break-off can be understood by analyzing the viscous drag on drop, if the effect of capillary instability is negligible. It was earlier discussed that the viscous drag scales as $\mu_1 V_c d_{drop}$ and the break-up occurs when this drag is comparable to surface tension force, $\sigma d_{tip}$. Note that $V_c$ is the velocity of fluid surrounding the drop when it is coming out of the constriction and not the average velocity ($V$) of the surrounding fluid which was used in calculating the capillary number as the surrounding fluid velocity increases in the constriction (due to reduction in area). The value of $d_{tip}$ is the thickness of the drop at the center of the constriction and is given by $d_{tip} = 2(R - d - f)$, where $R$ is the radius of the channel, $d$ is depth of the constriction, and $f$ is the film thickness. The film thickness is the thickness of the fluid surrounding the drop. The film thickness at the center of the constriction is generally very small compared to the other lengths. As the break-up starts taking place, the film thickness will increase, resulting in a lower surface tension force, which will promote break-up even more. Therefore, the surface tension force can be scaled by taking the maximum value of $d_{tip}$, which is $2\,(R-d)$. The break-up occurs if the viscous drag is more than the opposing surface tension force. When the drop starts coming out of the constriction, $V_c$ starts decreasing (due to increase in channel radius) and the diameter of the protruding drop starts increasing. Provided all other parameters are same, for a given increase in this protruding drop size, the shape parameter, $\bar{k}$, will influence if the break-up will take place as it will decide the rate of decrease of $V_c$. If there is sudden decrease in $V_c$, then the break-up will not take place due to insufficient viscous drag. For example, for a triangular constriction ($\bar{k}=0.5$), the increase in film thickness is very large as the drop comes out of the constriction, and hence the viscous drag is unable to promote break-up due to more prominent decrease in surrounding fluid velocity. For a rectangular constriction ($\bar{k}=0$), as the drop comes out of the constriction, there is a sudden increase in area, which decreases the surrounding fluid velocity to a great extent and the viscous drag is again not strong enough to ensure snap-off. Therefore, the break-up will take place for in-between values and it would indeed be interesting to find out how the shape parameter exactly influences this. Again, note that the capillary instability has been neglected here and it will surely come into picture if the capillary numbers are high or if jet like formations are taking place (for longer constriction lengths).

Table 1: Constriction area and shape parameter for different shapes.

| Constriction Shape | Area ($A_c$) | Shape Parameter ($\bar{k}$) |
|---|---|---|
| Rectangular | $Ld$ | $0$ |
| Circular | $\dfrac{\pi Ld}{4}$ | $1-\dfrac{\pi}{4}$ |
| Trapezoidal (with base length = $L$/3) | $\dfrac{2Ld}{3}$ | $\dfrac{1}{3}$ |
| Sinusoidal | $\dfrac{2Ld}{\pi}$ | $1-\dfrac{2}{\pi}$ |
| Triangular | $\dfrac{Ld}{2}$ | $\dfrac{1}{2}$ |
| Curved Triangular | $Ld(1-\dfrac{\pi}{4})$ | $\dfrac{\pi}{4}$ |

While closing this section, it is worth mentioning that in addition to surface shape, the effects of shape of the cross-sectional area have also been studied, which are not being discussed in detail in this paper. For example, Gauglitz et al. [48] studied bubble break-up in a square constricted capillary, whereas Ransohoff et al. [49] studied the effects of different types of noncircular geometries such as triangular, rectangular etc. on bubble snap-off. Kovscek and Radke [50] also studied gas bubble snap-off under pressure driven flow in noncircular capillaries.

### 3.2. Effect of constriction ratio:

The constriction ratio refers to the ratio of maximum diameter to the minimum diameter (also known as throat diameter) in the tube. It is also sometimes referred to as the constriction depth or the constriction gap. Since, it is a ratio of diameters, it can also be expressed in terms of radii. For a converging/diverging channel, the constriction ratio (C) is defined as:

$$C = \frac{R}{R-d} \qquad (3)$$

Tsai and Miksis [39] evaluated the bubble shapes keeping all other parameters same and varying only the constriction depth in a tube of radius 1 (all values have been reported in non-dimensional units with respect to channel radius) for $a = 1.2$, $\lambda = 0.001$, $Ca = 0.1$. They observed snap-off in a constricted channel of depth 0.6 ($C = 2.5$) whereas no such snap-off was observed in a constriction having a depth 0.5 ($C = 2$) as shown in Fig. 10. Thus, a higher constriction ratio implies a greater tendency of drop break-up. This is because a higher constriction ratio for the same radius means a lower value of $R - d$. From the force analysis performed earlier, it implies that the surface tension force acting now is lower (due to lower value of $d_{tip}$), hence making it easy for the viscous drag to break the drop.

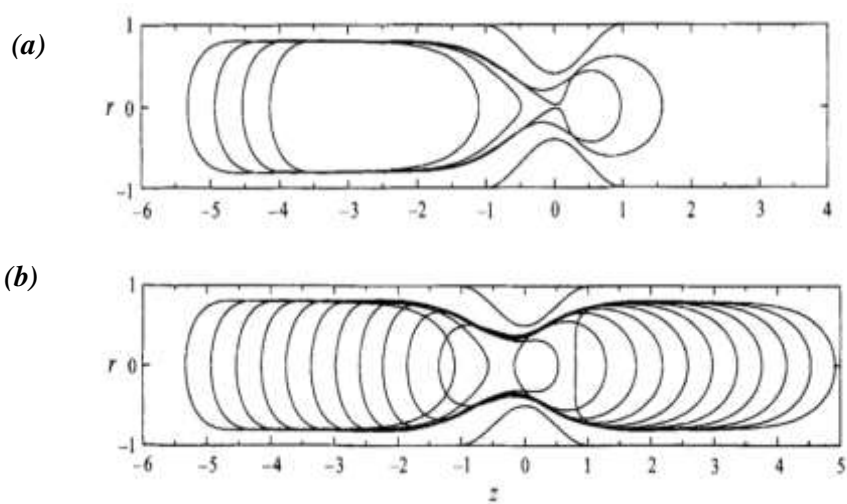

(a)

(b)

Figure 10: Effect of constriction ratio on snap-off of a bubble with $a = 1.2$, $\lambda = 0.001$, $Ca = 0.1$: (a) Constriction ratio = 2.5; (b) Constriction ratio = 2. Reproduced with permission from Journal of Fluid Mechanics 274, 197-217 (1994). Copyright 1994, Cambridge University press.

Martinez and Udell [51] used the boundary integral method to analyze flow through two different sinusoidal tubes having periodic constrictions with constriction ratios 3 and 1.8 respectively. There was a sudden decrease in additional pressure drop for the narrower tube ($C = 3$) as the drop approaches the constriction, which implies building up of higher pressures (or a higher resistance to drop flow) [Fig. 11 (c) and (d)].

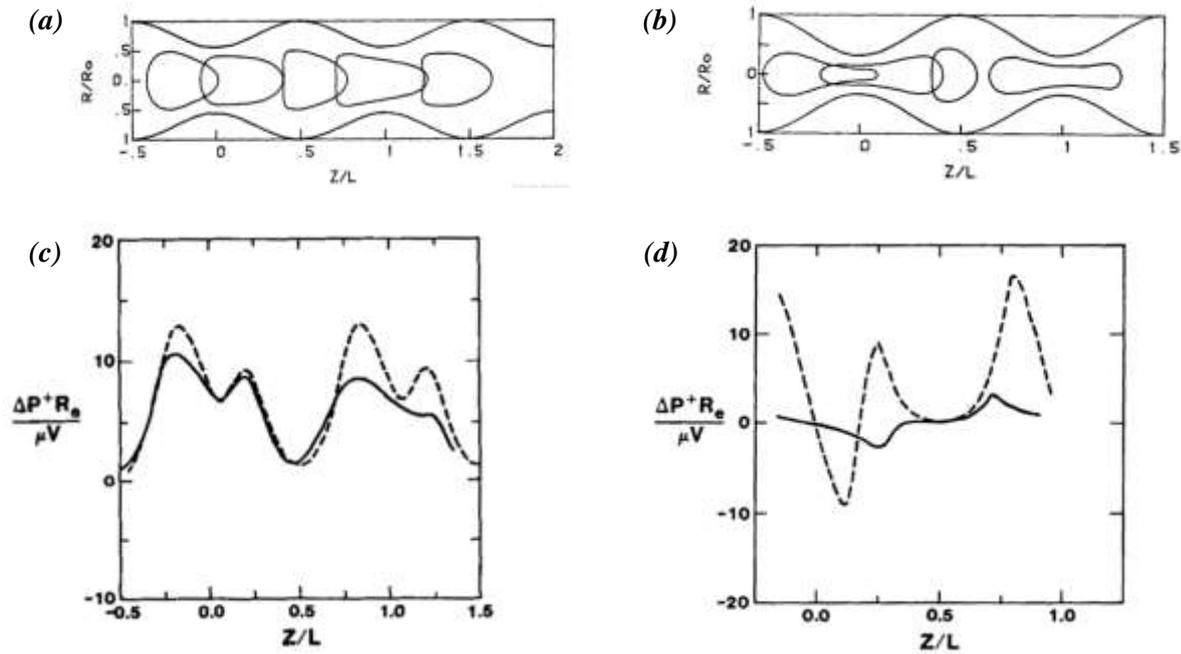

Figure 11: (a) Drop shape evolution in tube with constriction ratio 1.8, $a = 0.76$, $Ca = 0.5$, $\lambda = 10$; (b) Drop shape evolution in tube with constriction ratio 3, $a = 0.9$, $Ca = 0.5$, $\lambda = 1$; (c) Additional pressure drop in tube with constriction ratio 1.8; (solid line) $a = 0.76$, $Ca = 0.5$, $\lambda = 10$; (dashed line) $a = 0.76$, $Ca = 0.083$, $\lambda = 7.5$; (d) Additional pressure drop in tube with constriction ratio 3, $a = 0.90$, $\lambda = 1$, (solid line) $Ca = 0.5$, (dashed line) $Ca = 0.083$. Reproduced from M. J. Martinez and K. S. Udell, "Axisymmetric creeping motion of drops through a periodically constricted tube," AIP Conference Proceedings 197, 222 (1990), with the permission of AIP Publishing.

They also observed droplet deformation in the tubes though it cannot be compared directly as the other set of influencing parameters ($Ca, \lambda, a$) were different for the two cases. But it can be easily inferred that a greater constriction ratio implies more drop deformation [Fig. 11 (a) and (b)]. A decrease in drop speeds with increase in constriction ratio was also reported which can be attributed to higher resistances, though it is also a lot more dependent on drop size.

Nath et al. [9] also discussed the effect of constriction gap on various important parameters (Fig. 12). As mentioned in the previous section, they studied the behavioral changes in EMT cancer cells passing through rectangular microcapillaries having gaps of 10, 12 and 15 µm. For a constant cell size, the deformation index (ratio of drop length to its width) and the entry time increase with decreasing gap size due to enhanced deformations. Due to the same reason, the cell velocity in a broader gap is more. The deformation index increases with increasing cell size. As a greater cell size corresponds to more bulkiness, the entry time increases with the increase in cell size, whereas the transit velocity decreases.

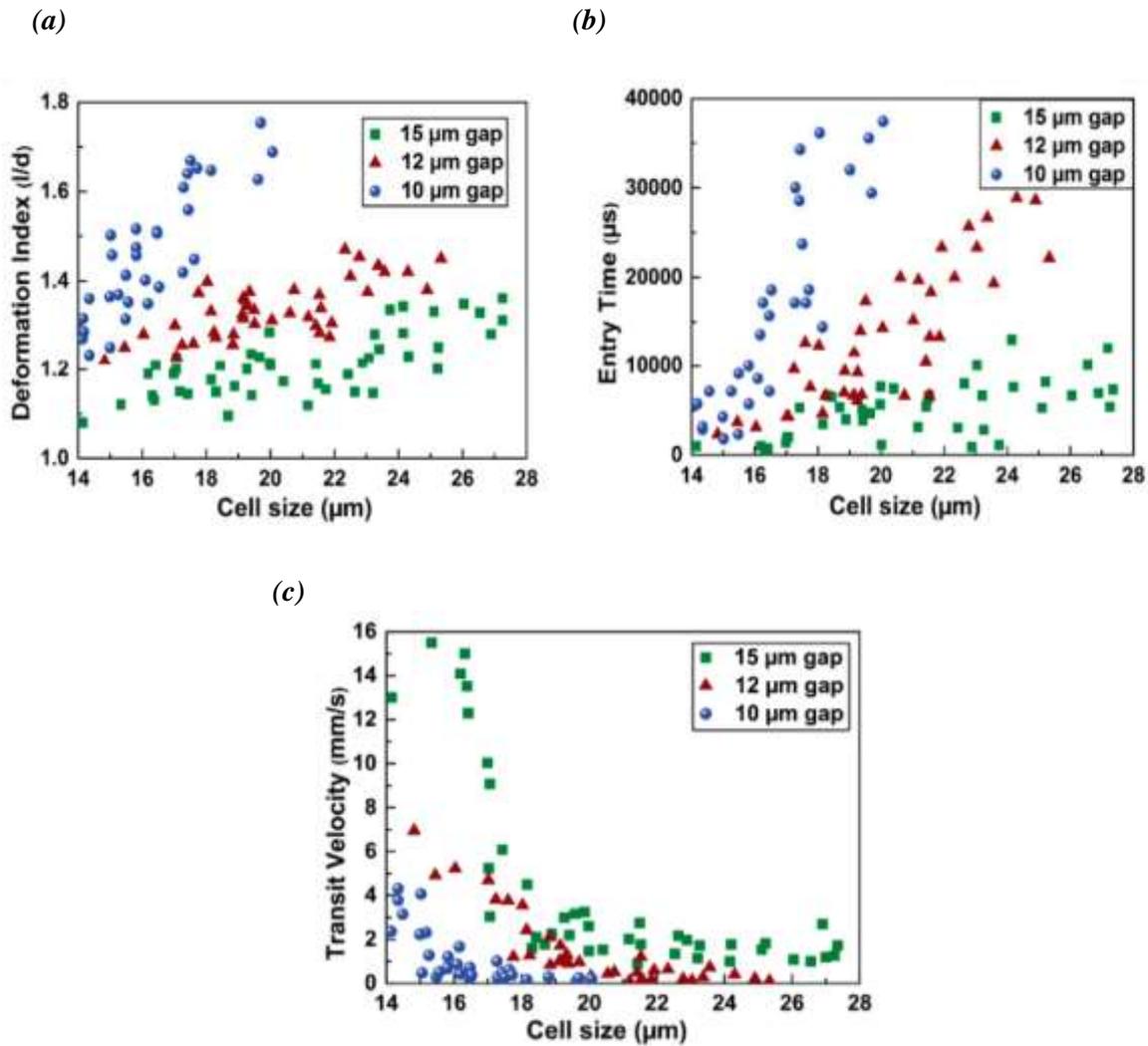

Figure 12: Effect of cell size and gap size on (a) Deformation index; (b) Entry time; (c) Transit velocity. Reproduced from J. Clin. Med. 8,1194 (2019). Copyright 2019, Multidisciplinary Digital Publishing Institute.

Ratcliffe et al. [52, 53] studied the effects of gravity on the squeezing of a deformable drop through a ring constriction. The circular ring can be thought of as a constriction with a semi-circular boundary. They observed that on increasing the drop-to-hole ratio ($a/b_s$), a transition occurs from squeezing to trapping [Fig. 13 (a)]. This means that instead of "squeezing" through the constriction, the drop starts getting "trapped" around the constriction for higher drop-to-hole ratios for a given Bond number (the Bond number is a non-dimensional number which gives the ratio of gravity forces to the capillary forces). This is because a lower drop-to-hole ratio implies a lower constriction ratio and hence lower pressures, which lead to a lower resistance to the drop flow. For the cases in which squeezing occurred, they found that the squeezing time scales as $T_s \sim (B - B_c)^{1/2}$, where $B$ is the Bond number and $B_c$ is the critical Bond number (the minimum bond number below which drop trapping occurs).

They also found out that on decreasing the ring cross-section area, a transition takes place from squeezing to trapping since a ring of lesser cross-section radius allows dripping of the drop on its outer surface. This can be seen in Fig. 13-(b): on decreasing the $a_s/b_s$ (cross-section-to-hole ratio) from 5 to 1 for same hole radius, a transition from squeezing to trapping occurs.

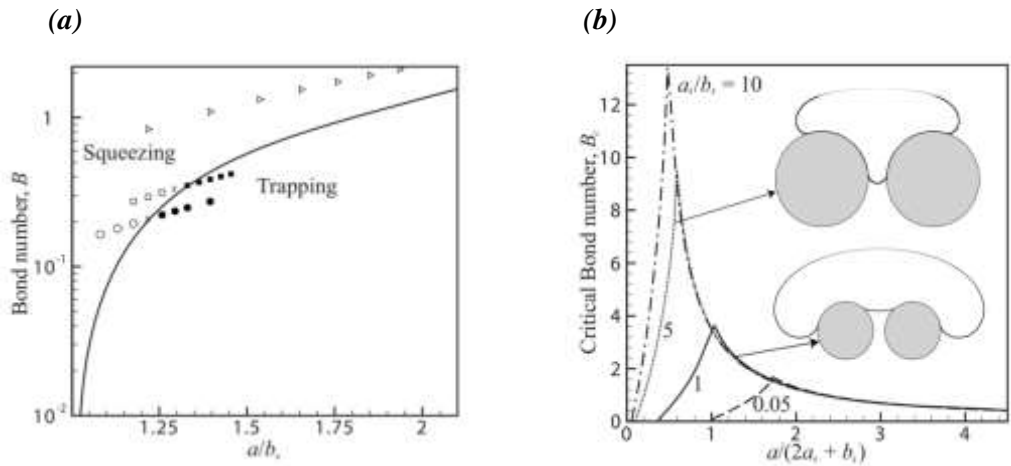

Figure 13: (a) Transition from squeezing to trapping by increasing the drop-to-hole ratio ($a/b_s$), for a fixed $a_s/b_s = 0.567$. The circles represent 100% water drops, the squares represent 75% water drops, the triangles are for 16% water drops, and the solid curve is the predicted critical Bond number; (b) Variation of critical bond number with cross-section radius ($a_s$), drop radius ($a$) and hole radius ($b_s$). Reproduced from T. Ratcliffe, A. Z. Zinchenko, and R. H. Davis, "Buoyancy-induced squeezing of a deformable drop through an axisymmetric ring constriction," Physics of Fluids 22, 082101 (2010), with the permission of AIP Publishing.

Similar results were observed by Bordoloi and Longmire [54], who studied the motion of gravity-driven deformable drops through a round edged (semi-circular) confining orifice. They observed that the drop gets captured for lower orifice-to-drop diameters ($d/D$) which correspond to lower constriction ratios, whereas it passes through the constriction for higher values of $d/D$ [Fig. 14]. Another interesting thing to note is that with an increase in Bond number, drop break-up starts taking place for same constriction ratios. This can again be explained by understanding the forces acting on the drop and treating the gravity forces analogous to viscous forces in this case. A higher bond number allows more deformation which allows more surrounding fluid to come in contact with drop. This promotes the film instability and hence leads to break-up.

Here, the Bond number is given as $Bo = \rho_c g D^2/\sigma$, where $\rho_c$ is the difference in densities of the two fluids and $g$ is the acceleration due to gravity.

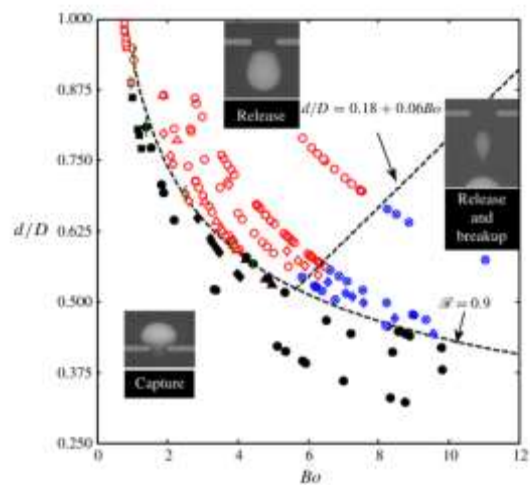

Figure 14: Regimes of drop capture, drop release and drop break-up for different orifice-to-drop diameters ($d/D$) and Bond numbers ($Bo$) ; circle: $\lambda = 0.15$; diamond: $\lambda = 0.67$; triangle: $\lambda = 1.81$; square: $\lambda = 0.02$. Reproduced with permission from Journal of Fluid Mechanics 759, 520-545 (2014). Copyright 2014, Cambridge University press.

Effect of constriction ratio was also analyzed numerically by Patel et al. [55], though their study is not in the low *Re* flow regime. Nevertheless, a good understanding is obtained on how the drop shape evolves with changing constriction ratio (Fig. 15). Their study also accounts for the effects of gravity by varying the Bond number (*Bo*). They also mentioned that the aspect ratio (ratio of bubble width to its length) increases for very high constriction ratios as the bubble starts elongating due to formation of low-pressure zones near its ends. Like other cases, the bubble velocity was found to decrease with increasing constriction ratio.

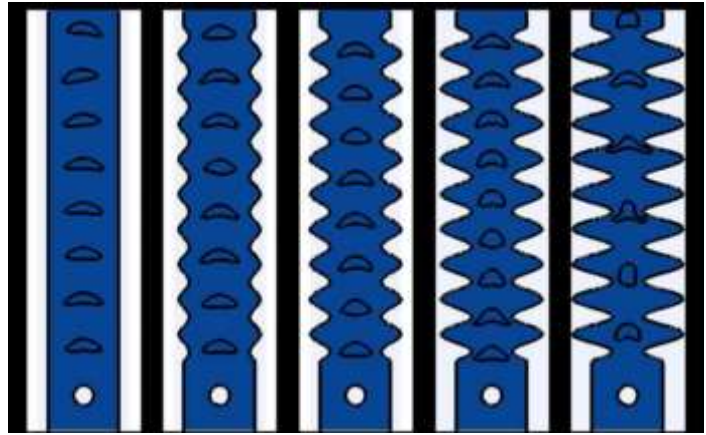

Figure 15: Evolution of drop shape in constricted channels with increasing constriction ratio from left to right at *Re* = 300 and *Bo* = 5. Reproduced from T. Patel, D. Patel, N. Thakkar, and Absar Lakdawala, "A numerical study on bubble dynamics in sinusoidal channels," Physics of Fluids 31, 052103 (2019). with the permission of AIP Publishing.

To conclude this section, it can be said that a greater constriction ratio corresponds to greater deformations, higher pressures, higher resistances and lower velocities (for larger drop sizes). A higher constriction ratio also implies a greater tendency for the droplet to snap-off. Thus, the constriction ratio can be modeled appropriately to promote phenomenon of drop break-up. Interesting shapes involving elongation are also observed for narrower channels.

### 3.3. Effect of phase angle between constrictions:
Phase angle refers to shift in phase between the periodic constrictions at the upper-wall and the lower-wall. The cases analyzed above have zero phase angles. The phase angle can vary from 0º to 360º. A phase angle of 360º implies a shift by one full constriction length and hence is the same as 0º. Having non-zero phase angles can change the flow parameters significantly. It can also have a considerable effect on the drop shape as shown below.

Sauzade and Cubaud [56] studied the effect of phase angle on the bubble shapes. The end of the upper constriction points to the middle of the lower constriction, thus indicating a phase angle of 180º. The constriction in this case resembles a triangular shape, though there is some curvature too. It can be observed that bubbles start acquiring velocities in vertical direction as well as asymmetrical shapes with respect to the horizontal axis (Fig. 16). Bubbles also seem to attain quite intriguing shapes, such as pointed triangles and zigzag quadrilaterals.

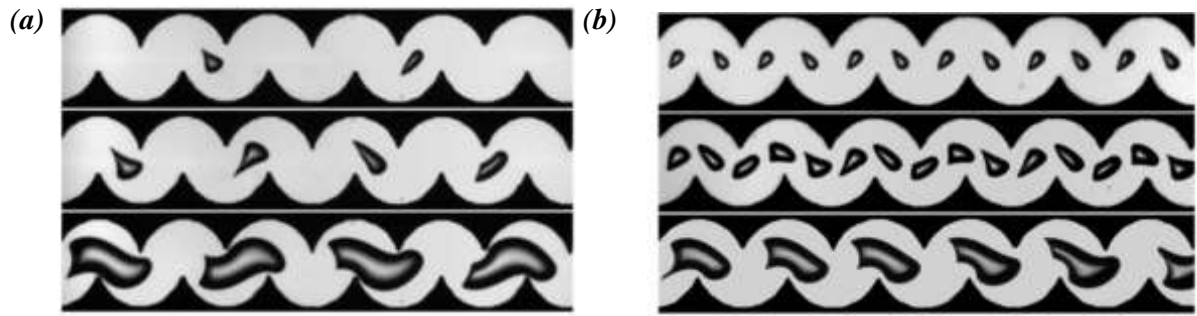

Figure 16: Effect of non-zero phase angle on bubble shapes in highly viscous oils with increasing gas volume fraction from top to bottom, (a) Air bubbles in 1,000 cS oil; (b) Carbon dioxide bubbles in 10,000 cS oil. Reproduced from M. Sauzade and T. Cubaud, "Bubbles in complex microgeometries at large capillary numbers," Phys. Fluids 26, 091109 (2014), with the permission of AIP Publishing.

Though the velocities inside the bubbles have not been computed, this type of motion hints that this type of geometry may enhance mixing of liquids inside the confined bubbles. Mixing requires generation of vortices within the complete bubble so that various chemical and biological reactions can take place. In flat channels having zero phase, efficient mixing is not possible since the vortices are formed in separate halves of the bubbles. Though these vortices may mix the contents of each half, there is no fluid exchange between the two halves. Thus, as of now, special zigzag, serpentine-like, rectangular corners and other similar channels are employed to enhance mixing [10, 57-60]. If it would be possible to ensure mixing just by altering the phase angle, it would indeed form a very fascinating application to be explored further.

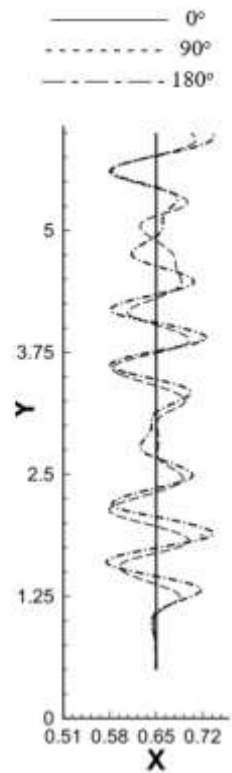

Figure 17: Effect of phase-angles on bubble centroid path in an inertial flow with $Re = 300$ and $Bo = 5$, for phase angles (a) 0°, 90° and 180°. Reproduced from T. Patel, D. Patel, N. Thakkar, and Absar Lakdawala, "A numerical study on bubble dynamics in sinusoidal channels," Physics of Fluids 31, 052103 (2019). with the permission of AIP Publishing.

Patel et al. [55] also observed the effect of phase angle in an inertial flow. They found that in constrictions with non-zero phase angles, the bubble centroid undergoes considerable deviation from its straight path which was observed for a zero phase angle constriction (Fig. 17). They also found that changing phase angle also changes the bubble velocities and the aspect ratios. The maximum velocity as well as the velocity fluctuations decrease with increasing phase angle. The maximum values of aspect ratio increased with increasing phase angle, while the fluctuations in the same were found to show a decrease. Also, the phase angles in their paper have been reported with a phase shift of 180$^o$ from how it has been defined in this paper.

### 3.4. Effect of length of the constriction

Tsai and Miksis [39] computed the effects of length of constriction on the bubble flow (Fig. 18). They took three different values of constriction lengths- 1.5, 2 and 4 (all values in non-dimensional units with respect to channel radius). While snap-off was observed in all the three cases, there was significant difference in snap-off parameters, such as the snap-off time (the time taken for the bubble to break-up) and the snap-off bubble radius (the radius of the front bubble after snap-off), which have been tabulated in Table 2.

Table 2: Effect of constriction length on snap-off time and bubble radius after snapping-off. The constriction length and bubble radius have been non-dimensionalized with length scale, $R$, whereas the snap-off time has been non-dimensionalized by the timescale, $R/V$ (Values obtained from Tsai and Miksis [39]).

| Constriction Length ($l$) | Time taken to snap-off ($\tau$) | Bubble Radius ($b_r$) |
|---|---|---|
| 1.5 | 0.68 | 0.77 |
| 2 | 0.42 | 0.64 |
| 4 | 0.45 | 0.68 |

Both snap-off time and bubble radius seem to follow a decreasing-increasing trend with increasing constriction length, implying that there may be minima somewhere in between. But, since the number of observations is too less, nothing concrete can be established, and this remains another area which can be studied further to get meaningful insights into this multiphase problem. Also, for longer lengths, the capillary instability will also start playing an important role since there will be formation of jetting regime. Therefore, it will be the combined effect of viscous drag and capillary mechanisms which will decide the drop shape, break-up and size.

### 3.5. Effect of periodicity and spacing between the constrictions

This problem has not been studied quite thoroughly, but a few inferences can certainly be made based on some observations. Most of the zero phase angle cases that have been seen above only consider a single constriction. Having multiple constrictions, even in a zero phase angle case, can lead to interesting results. Sauzade and Cubaud [56] studied the motion of drops in corrugated channels with semi-circular periodic constrictions. They observed many interesting drop shapes varying throughout the geometry and periodic in nature (Fig. 18). When the flow velocity of the bubble is increased, the bubbles are seen to deform from out of phase to in-phase mode. In the out of phase mode, the bubble is less deformed and shows periodicity in shape after every third periodic constriction whereas for in-phase mode, the bubble is elongated and shows periodicity in shape after every alternate constriction. Martinez and Udell [51] also observed different periodic shapes in periodic constriction channels (Fig. 11 in section 3.2).

The parameter which will play a major role in periodic constrictions having a fixed shape and length would be the spacing between the constrictions. For the above cases, the spacing between two consecutive constrictions was negligible as the constrictions almost touched each other. A single constriction can be visualized as a constriction with the spacing tending to infinity. Having some fixed spacing and manipulating flow within it can lead to interesting phenomena like drop coalescence, as shown next.

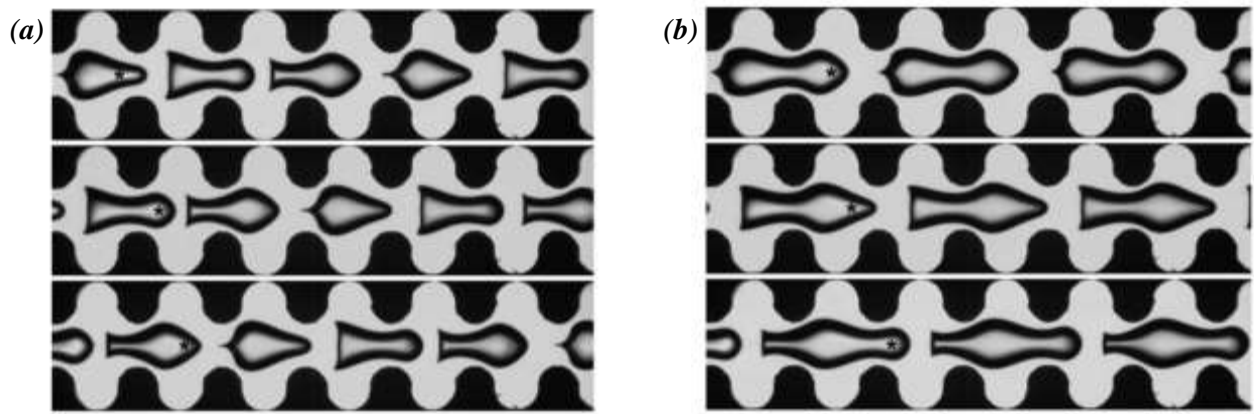

Figure 18: Time-series of elongated air bubbles flowing with an oil of viscosity $10^3$ cS in a mildly deformed microchannel. Different initial liquid-gas flow conditions can lead to different phase shifts: observe that bubbles in (a) are out of phase (lower velocity), whereas bubbles in (b) are in-phase (higher velocity). Reproduced from M. Sauzade and T. Cubaud, "Bubbles in complex microgeometries at large capillary numbers," Phys. Fluids 26, 091109 (2014), with the permission of AIP Publishing.

Seemann et al. [10] in their review on droplet microfluidics summarized how droplet coalescence can take place by various mechanisms. They also emphasized the use of channel geometry and showed how different types of geometries can lead to drop coalescence. In [Fig. 19 (a)], the drop has been stopped by enlarging the channel and providing appropriate spacing, thus allowing the coming drop to merge into it. In [Fig. 19 (b)], adequate spacing has been provided between the constrictions, which allows the continuous phase to flow around the droplet and slow it down. This allows the succeeding drop to merge into the first one. This type of drop coalescence by varying the geometry is indeed a very fascinating application and has also been studied by many researchers [61-65].

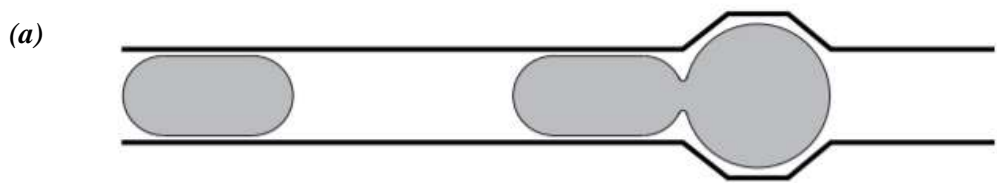

*(b)*

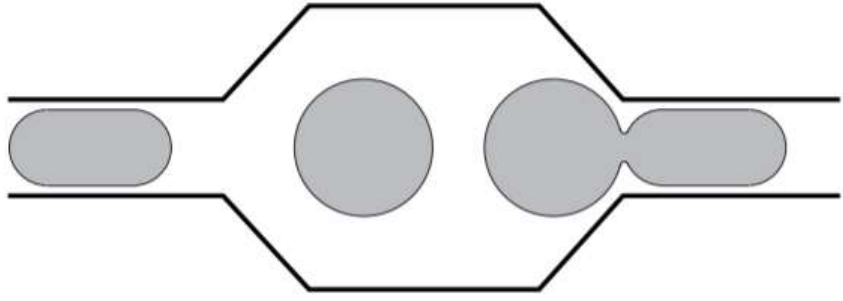

Figure 19: Drop coalescence by (a) stopping the droplet in between the spacing between the two constrictions; (b) slowing down the droplet and allowing the succeeding droplet to merge into it. Reproduced with permission from Rep. Prog. Phys. 75, 016601 (2012). Copyright 2012, IOP Publishing.

Thus, it can be inferred that periodicity and spacing not only lead to interesting drop shapes in the flow but also play a major role in the phenomenon of coalescence.

## 4. Scaling analysis for multiphase flow through constricted passage

From the above review of different studies, it is found that a more systematic study of flow through a constricted passage is necessary to explore the various application of drop/bubble deformation. A common criterion for droplet pinch-off through constricted passages with different surface topography is still missing in literature. A scaling analysis is proposed here to model such flows. We plan to numerically simulate the phenomena of droplet deformation and propagation through passage with varying surface topography in future. In literature, the common numerical methods used are boundary integral method [33, 34, 39, 44, 46, 53], VOF method [66-72], dual-grid level set method (DGLSM) in the context of an Immersed Boundary (IB) [55] and level set [47] methods. The interface capturing methods of coupled level set and volume of fluid [68-72] have also been used for investigating droplet deformation and pinch-off due to impact on fluid surface. The authors believe with the present formulation of the problem, the unique criteria for drop deformation through constricted passage independent of surface geometry can be found.

We aim to list the various non-dimensional parameters involved in the process. The main governing forces in the flow are the viscous forces, interfacial forces and the gravity (only if density difference is present). It has already been seen that the pressure drop, flow speed, drop shape and multiphase phenomena depend upon a lot of parameters. A schematic of typical periodic constriction is shown in Fig. 20 with different length scales labelled. The main parameters that have been identified are the drop-radius ($A$), mean Poiseuille flow velocity ($V$), radius of the channel ($R$), surface tension ($\sigma$), viscosity of surrounding fluid ($\mu_1$), viscosity of the drop ($\mu_2$), density difference between the two fluids ($\rho_c = \rho_2 - \rho_1$), the shape parameter ($k$), depth of the constriction ($d$), length of the constriction ($L$), spacing between two consecutive constrictions ($S$), and the phase shift ($P$). Phase shift is the distance between the starting of the upper constriction and the corresponding lower constriction. This parameter will give an idea of the phase angle.

Therefore, in total, there are twelve-dimensional parameters having units consisting of three independent fundamental physical quantities - length, mass and time. Thus, according to Buckingham $\pi$ Theorem, nine non-dimensional numbers are obtained, namely, capillary Number, $Ca = \mu_1 V/\sigma$, Bond Number, $Bo = \rho_c g R^2/\sigma$, viscosity Ratio, $\lambda = \mu_2/\mu_1$, spacing parameter, $\bar{S} = S/R$, length parameter, $\bar{L} = L/R$, constriction gap parameter, $\bar{D} = d/R$, drop size, $\bar{A} = A/R$, the phase shift, $\bar{P} = P/R$, and the shape parameter, $\bar{k} = 1 - A_c/Ld$.

Thus, the non-dimensional pressure drop ($\Delta PR/\mu_1 V$), non-dimensional velocity ($u/V$) and the drop shape are a function, $f(Ca, Bo, \lambda, \bar{S}, \bar{L}, \bar{D}, \bar{A}, \bar{P}, \bar{k})$, of these nine quantities.

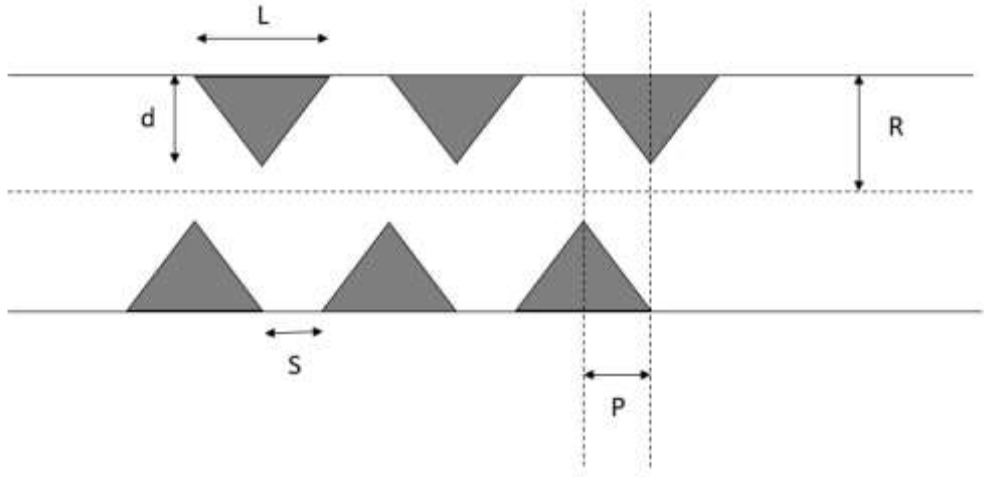

Figure 20: Schematic depicting the geometry parameters in periodic triangular constrictions.

## 5. Applications

Multiphase flows through a constricted channel find many diverse applications. Fig. 21 lists some of these applications. Starting with the medical applications, multiphase flows through a constricted microchannel have been utilized in 3D bioprinting [73]. Artificial tissues and organs are now being made using bioprinting which uses concept of microfluidics for cell interaction and deposition. The blood in human body flow through various microcapillaries and understanding this flow in detail can help find the cause a well as the cure of many diseases. The cells can be modeled as one phase and the blood as another phase. We have already seen this in a study on cancer cells by Nath et al. [9]. Blood flow through zigzag and hyperbolic channels are also being studied to determine the mechanical properties of fibrin networks [10]. Multiphase flows through constriction are also used in various biological reactions. Shang et al. have discussed the application of droplet microfluidics in small-molecule detection and biological macromolecule analysis [12]. Nowadays, fuel cells are being used as energy sources owing to their high efficiency, smaller size, and environment friendly nature. They also consist of various phases flowing through constricted microchannels [74]. Micro-electro-mechanical-systems (MEMS) have become quite popular nowadays and are find applications everywhere around us, from automotive to our devices like mobiles and printers, from pressure sensors to biotechnology. This is because of the many advantages they offer including small size, low power consumption, low cost, etc. Many of these systems use the concepts of microfluidics and involve multiphase flows through microchannels [75]. We had earlier said that flow through a porous media can be approximated very nicely using the model of drop flow through a constriction. Many industries employ the flow of emulsions through porous media. For example, Khan et al. used multiphase porous media modeling to develop a physics-based model of food processing, aimed to optimize energy consumption and improve food quality [8].

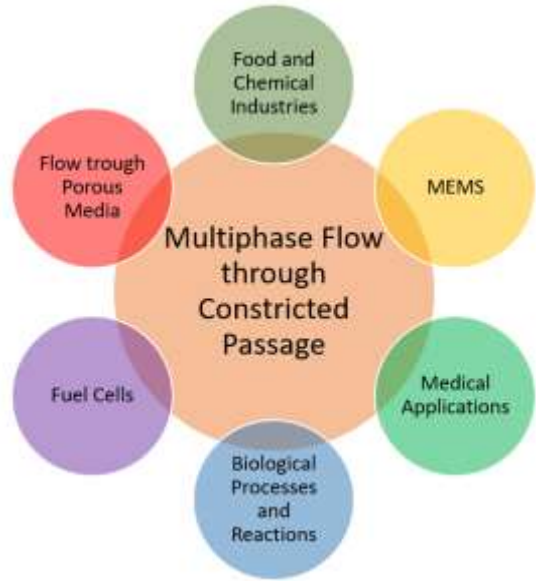

Figure 21: Applications of multiphase flow through a constricted passage.

## 6. Summary and Discussion

In this paper, we have studied the recent progress regarding the drop/ bubble morphology as it flows through constricted passages and tried to understand how the surface topography affects the drop flow. Since different studies have taken different parameters, for most of the cases it was difficult to analyze how changing only a particular geometry parameter affects the flow. Nonetheless, we obtained a general idea of how different parameters affect the drop morphology and saw how some parameters can help in promoting certain physical phenomena which can find many important applications. We also made an attempt to understand the effects of different geometric parameters through the aid of the physics and mechanisms governing this problem.

Drops/bubble deformation, velocity and pressure gradient across the straight microchannel are dependent on the droplet size, viscosity ratio and the capillary number. Reviews show that the primary forces involved in such flows are the viscous forces and the interfacial forces, which tend to deform the drop/bubble surface.

Analysis of surface topography is done based on constriction shape, constriction ratio, phase angle between the constrictions, length of the constriction and constriction periodicity. Having multiple constrictions with some fixed spacing and manipulating flow within these can lead to phenomena like drop coalescence and mixing. In case of wavy channel, in certain circumstances, constrictions could lead to very high pressures which may lead to drop being retained at certain locations in the channel (trapping). To avoid such situation, systematic investigation of how each geometric parameter specifically changes the flow behavior and drop shape, when other parameters are kept constant, is necessary.

Different shapes have different effects on the drop in terms of deformation and velocities. Certain shapes also promote drop break-up in comparison to other shapes. For a greater constriction ratio, snap-off of the bubble is observed due to lower surface tension force compared to the viscous drag. Increase in length of the constriction may lead to capillary instability coming into picture and hence it needs to be evaluated how it affects the drop break-up and size. Introducing phase angle gives rise to irregular drop shapes and it would also be interesting study to explore the possibility of mixing with a change in phase angle for constrictions of different shapes. Periodicity and spacing not only lead to interesting drop shapes in the flow but also play a major role in the phenomenon of coalescence. 3D modeling of drop morphology through constricted passages by using the scaling non-dimensional parameters mentioned in the paper will enhance the understanding of the problem. The vast applications of flow through porous structure, blood flow, mixing can be explored further.

## Data Availability

Data sharing is not applicable to this article as no new data were created or analyzed in this study.